# Enabling magnetic resonance imaging of hollow-core microstructured optical fibers via nanocomposite coating

ROMAN E. NOSKOV,[1,6] ANASTASIA A. ZANISHEVSKAYA,[2,3] ANDREY A. SHUVALOV,[2,3] SERGEI V. GERMAN,[4] OLGA A. INOZEMTSEVA,[3] TARAS P. KOCHERGIN,[3] EKATERINA N. LAZAREVA,[3,5] VALERY V. TUCHIN,[3] PAVEL GINZBURG,[1] JULIA S. SKIBINA,[2,3] AND DMITRY A. GORIN[4,7]

[1]*Department of Electrical Engineering, Tel Aviv University, Ramat Aviv, Tel Aviv 69978, Israel*
[2]*SPE LLC Nanostructured Glass Technology, Saratov 410033, Russia*
[3]*Saratov State University, Saratov 410012, Russia*
[4]*Skolkovo Institute of Science and Technology, Moscow 143026, Russia*
[5]*Tomsk State University, Tomsk 634050, Russia*
[6]*nanometa@gmail.com*
[7]*D.Gorin@skoltech.ru*

**Abstract:** Optical fibers are widely used in bioimaging systems as flexible endoscopes that are capable of low-invasive penetration inside hollow tissue cavities. Here, we report on the technique that allows magnetic resonance imaging (MRI) of hollow-core microstructured fibers (HC-MFs), which paves the way for combing MRI and optical bioimaging. Our approach is based on layer-by-layer assembly of oppositely charged polyelectrolytes and magnetite nanoparticles on the inner core surface of HC-MFs. Incorporation of magnetite nanoparticles into polyelectrolyte layers renders HC-MFs visible for MRI and induces the red-shift in their transmission spectra. Specifically, the transmission shifts up to 60 nm have been revealed for the several-layers composite coating, along with the high-quality contrast of HC-MFs in MRI scans. Our results shed light on marrying fiber-based endoscopy with MRI to open novel possibilities for minimally invasive clinical diagnostics and surgical procedures *in vivo*.

## 1. Introduction

Optical bioimaging via using fibre-optic endoscopes has supplied a large number of opportunities for retrieving information from remote and delicate places enabling accurate visualization of pathologies, neural activity, tissue structure and many others. Specifically, optical endoscopy is considered as a promising tool for monitoring, diagnosis and detection of diseases in luminal organs such as the gastrointestinal tract [1], the coronary arteries [2], and the pulmonary airways [3]. To be used in organs having natural openings, flexible endoscopes are typically made of optical fibre bundles [4] that consist of up to ~100,000 individual fibers in a closely packed arrangement. Such instruments, also known as fibrescopes, collect light by a miniaturized objective [5], a gradient-index (GRIN) lens [6] or a metalens [7]. The image is created on an input facet of a coherent fibre bundle and guided by it.

As an auspicious alternative allowing minimally invasive penetration into tissue, a single multimode fiber operating lens-free on principles of digital holography was offered [8]. This system is especially attractive for high-resolution observations of neuronal activity *in vivo* inside deep brain areas [9]. However, currently the precise position of the fiber endoscope after penetration into the brain can be defined post-mortem only. At the same time, the study of so complex system as a brain *in vivo* whose functioning is still mainly unknown requires extremely exact spatial operating with endoscopic probes in real time. Such opportunity can be provided by magnetic resonance imaging (MRI). Thin optical fibers, though, have a vanishing MRI contrast.

Here, we develop, for the first time to our knowledge, a simple and cheap method of making hollow-core microstructured fibers (HC-MFs) visible in MRI. Our approach is based on Layer-by-Layer assembly (LbLA) of oppositely charged polyelectrolytes and magnetite nanoparticles (MNPs) on the inner core surface of HC-MFs. Originally, LbLA was discovered by Iler in 1966 for oppositely charged inorganic nanoparticles [10]. Later Decher and Hong have expanded this approach on the polyelectrolytes [11]. LbLA is realized by the "bottom-up" principle, allowing accurate variations in the thickness and the chemical composition of the coating as well as the volume fraction of nanoparticles. In particular, this technique has been applied to cover glass, quarz and silicon substrates by composite multilayers containing polyelectrolytes and MNPs [10–14]. Also LbLA has been used to fill capillaries of HC-MFs with different materials, such as polymers [15] and metals [16], to reach specific functions [17–22].

As contrast agents for MRI, we chose magnetite ($Fe_3O_4$) nanoparticles which are fully compatible with our technique of LbLA and have a strong magnetic susceptibility. The latter, in particular, provides a large signal-to-noise ratio, allowing for visibility of HC-MFs in low-magnetic-field MRI scanners. This opens promising perspectives for using functionalized HC-MFs in MRI-controlled endoscopic surgery *in vivo* [23]. Additionally, the nanocomposite coating leads to the red-shift in the transmission spectrum of the HC-MF that can be used as an additional degree of freedom for the fiber post-processing tunability. We analyze the impact of the coating thickness on the performance of the system and find the optimal number of the composite layers resulting in artifact-free and high-quality contrast of HC-MFs in MRI.

## 2. Results and discussion

### 2.1 Preparation of samples

We consider the HC-MF containing five functional concentric layers of capillaries and the external buffer layer (Fig. 1 (a)) drawn from the custom-made soft glass with the refractive index $n = 1.519$ at the wavelength 550 nm (the dependence of the glass refractive index on the wavelength is presented in Appendix A) [24]. The diameter of the central capillary is 194 μm, the thickness of the walls for 1, 2, 3, 4 and 5th layer of capillaries are 1.8, 2.6, 3.3, 4.1, and 4.7 μm, the diameters of the capillaries belonging to 1, 2, 3, 4 and 5th layer are 14, 19, 25, 31, and 35 μm, respectively. Spectral properties for this kind of HC-MFs were discussed in detail in [25–27].

Figure 1 shows the HC-MF and the scheme of the coating formation. We start with hermetic coupling of the HC-MF to the pipette dispenser tip which is used to fill the fiber with polyelectrolytes and MNPs. The structure is washed with the deionized water twice (as shown in Fig. 1(b)). Then the LbLA is employed for fabrication of the interior coating. Firstly, the HC-MF is filled by poly(diallyldimethylammonium chloride) (PDDA) solution, carrying positive charge (while fiber glass is negatively-charged). The solution is kept inside the HC-MF for 10 min, and after that the sample is drained. PDDA molecules are adsorbed onto the interior surface due to Coulomb interaction with the glass surface. Then the structure is washed 3 times with deionized water. Next, positively-charged MNPs are terminated to the PDDA layer by the same means. The initial magnetite hydrosol has concentration of 15 mg/ml and is diluted with deionized water (a ratio 1/10 (v/v)). The procedure is repeated to deposit the desired number of bilayers of PDDA and MNPs. All in all, we have prepared 15 samples with 1, 2, 3, 4 and 5 nanocomposite bilayers of different molecular weight (MW) for PDDA (low, medium, high).

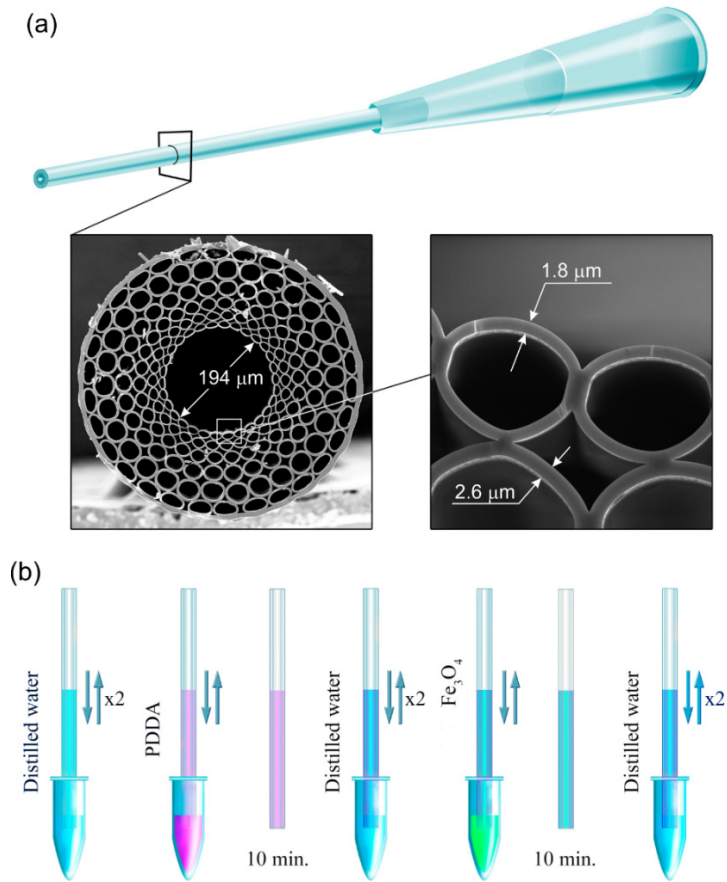

**Fig. 1.** Schematic showing preparation of the samples. (a) Hermetic coupling of the HC-MF to the pipette tip. The scanning electron microscopy (SEM) images denote the cross-section of the HC-MF. (b) The scheme of the PDDA/MNPs coatings formation. At first, the structure is washed twice by the deionized water. Then, polyelectrolytes and magnetite nanoparticles are adsorbed alternately onto the interior surface of the HC-MF by feeling the fiber with the polyelectrolyte solution for 10 min followed by washing with the deionized water after each adsorption step. The procedure is repeated to depose the desired number of bilayers.

## 2.2 Magnetic resonance imaging of HC-MFs with nanocomposite coating

The samples were studied by MRI clinical scanner Philips Achieva 1.5 T. T1- and T2-weighted quick "spin-echo" (turbo spin-echo (TSE)) and T1 weighted "Fast Field Echo" (is equal to the "Gradient Echo") protocols were applied. MNPs mainly reduce the transverse relaxation time T2, yielding the darker staining of corresponding marked areas. HC-MFs with 1, 3 and 5 nanocomposite bilayers were placed into microcentrifuge tubes filled with water (Fig. 2(a)). Figure 2(b) shows the 3D reconstruction of the samples obtained from T2 weighted scans. Increasing the number of bilayers leads to better contrast in the region of interest (compare the samples with 1, 3, and 5 bilayers in Figs. 2(b) - (d)). However, the 5-bilayer coating already gives rise to artifacts as a result of a too strong magnetic response. This is evident in increasing the fiber thickness in magnetic resonance (MR) scans (see Figs. 2(b) - (d)) while the physical thickness for all samples is the same.

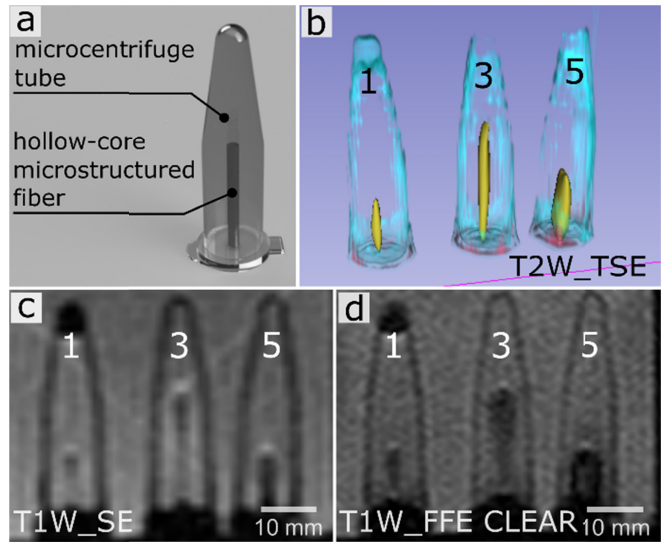

**Fig. 2.** Magnetic resonance investigation of HC-MFs coated by nanocomposite bilayers. (a) Schematic of a sample: HC-MF is placed in the microcentrifuge tube filled with water. (b) 3D reconstruction obtained by T2 weighted (turbo spin-echo) MR scanning. HC-MFs are marked by yellow. (c) T1 weighted (spin-echo) MR image in the longitudinal plane. (d) T1 weighted (fast field echo) MR image in the longitudinal plane. The numbers '1', '3′, and '5′ denote the number of nanocomposite bilayers in the sample. The lengths of the samples: '1' – 0.9 cm; '3′ – 1.65 cm and '5′ – 0.8 cm.

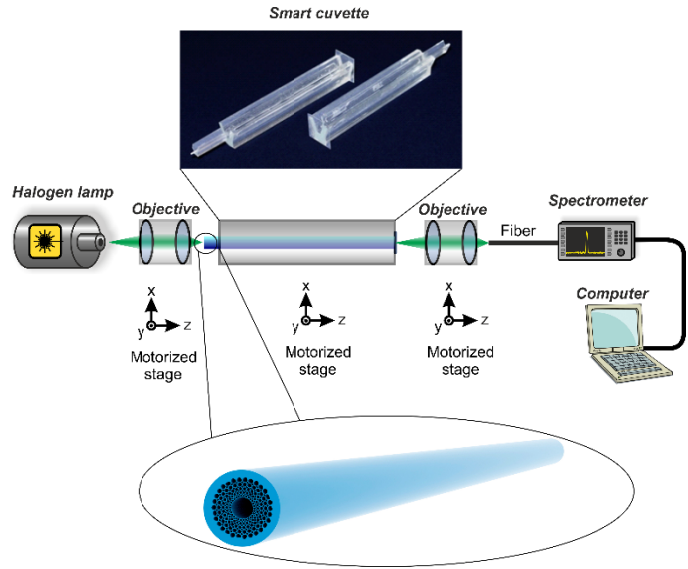

**Fig. 3.** Schematics of the experimental setup for transmission characterization of HC-MF samples. The illumination of the broadband light source (halogen lamp) is launched into the HC-MF via an objective. The HC-MF is integrated in a smart cuvette (61-mm length). The input and output objectives and the cuvette are adjusted with three-axis motorized stages. The output spectrum is measured by the spectrometer Ocean Optics HR4000. The resulted data are analyzed by a personal computer.

## 2.3 Impact of nanocomposite coating on transmission of HC-MFs

To characterize optical transmission of the samples, we use the experimental setup, schematically shown in Fig. 3. The HC-MF is placed into a specially designed smart cuvette and fixed on a three-axis movable platform. In addition to fiber holding, the smart cuvette also can control meniscus formation that is important for light coupling when the fiber is filled with liquid. This type of cuvette has been served as a basic replaceable element of a biosensor for blood typing [28].

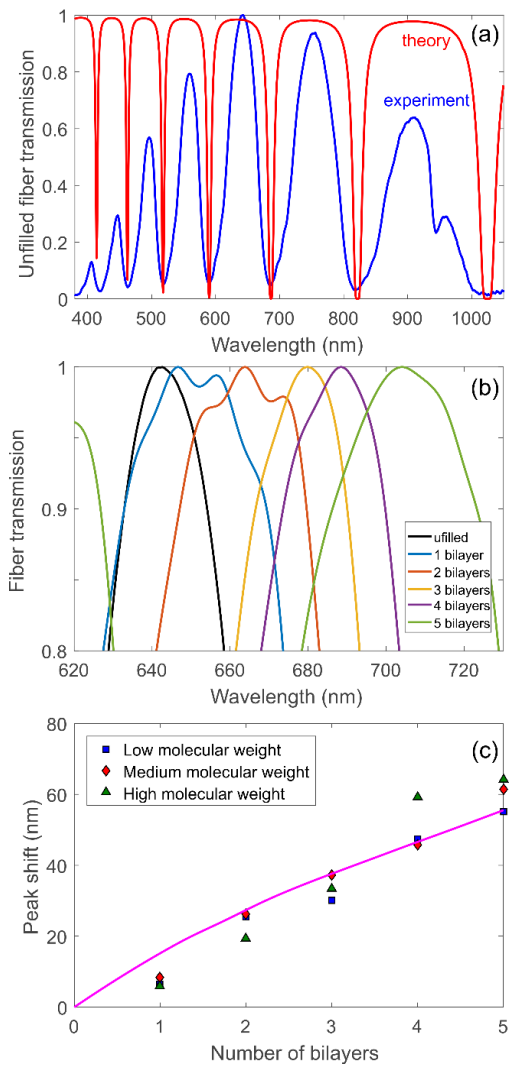

**Fig. 4.** Characterization of optical transmission for HC-MFs. (a) The transmission spectrum of an unfilled HC-MF obtained by measurements (blue) and theoretical calculations (red). (b) Modifications in the selected HC-MF transmission window induced by coating with different numbers of bilayers (a high molecular weight) in comparison with the unfilled sample. (c) The transmission maximum as a function of the number of nanocomposite bilayers. The purple line denotes a theoretical prediction. The thickness of every nanocomposite bilayer is assumed to be 25 nm. The theoretical calculations have been performed for the TE-polarized light.

Microscope objectives (40´) are used to launch the white light from a halogen lamp into the fiber and collect the outcoming radiation. The spectrum of the output signal is measured by the spectrometer Ocean Optics HR4000 operating in the visible/NIR range.

Figure 4 (a) shows the transmission spectra for the unfilled HC-MF. We adjust the system to provide the maximal coupling efficiency at the wavelength 640 nm. As a result, it decreases towards the edges of the spectral range, leading to comparative reduction of the transmission peaks. Deposition of the nanocomposite coating gives rise to the red shift of the transmission windows, reaching 60 nm for the 5-layers coating (Figs. 4 (b) and (c)). The similar results are obtained for low, middle and high MW.

Next, we present the theoretical background to calculate the transmission for HC-MFs and predict the shift in the transmission windows induced by nanocomposite bilayers. The light guidance in HC-MFs is provided by the resonant light reflection from the cladding layers, which can be characterized in terms of the model describing Fabry-Perot resonator [29], [30]. For the sake of simplicity, the HC-MF cladding is considered as concentric cylindrical layers with air, glass and nanocomposite refractive indexes. Within the geometric optics treatment, one can calculate the fiber transmission spectrum as follows [30]

$$T = R^m, \qquad (1)$$

where $R$ is the coefficient of reflection from the cladding and $m$ is the number of such reflections for the light ray, propagating in the fiber core, on the fiber length $L$. The condition $R = 1$ corresponds to the anti-resonance between the Fabry-Perot modes of the cladding and the modes of HC-MF propagating in the core. Assuming the paraxial approximation, one can derive for the fundamental fiber mode [30]

$$m = \frac{\lambda L}{2 d_0^2 n_0}, \qquad (2)$$

where $d_0$ is the core diameter, $\lambda$ is the wavelength, and $n_0$ is the refractive index of the fiber core filling (i.e., air in our case).

Accounting for an irregular transverse structure of the considered HC-MF (Fig. 1 (a)), we follow to P. Yeh [31] and employ the general transfer-matrix method to calculate $R$, which can be presented as

$$R = \left| \frac{M_{21}}{M_{11}} \right|^2. \qquad (3)$$

Here $M_{11}$ and $M_{21}$ are the elements of the transfer-matrix that is given by

$$\mathbf{M} = \begin{pmatrix} M_{11} & M_{12} \\ M_{21} & M_{22} \end{pmatrix} = D_0^{-1} \left[ \prod_{l=1}^{N} D_l P_l D_l^{-1} \right] D_s,$$

where $N$ is the number of layers,

$$D_l = \begin{cases} \begin{pmatrix} 1 & 1 \\ n_l \cos \vartheta_l & -n_l \cos \vartheta_l \end{pmatrix} \text{ for TE polarization,} \\ \begin{pmatrix} \cos \vartheta_l & \cos \vartheta_l \\ n_l & -n_l \end{pmatrix} \text{ for TM polarization,} \end{cases}$$

$$P_l = \begin{pmatrix} \exp(i 2 \pi n_l d_l \cos \vartheta_l / \lambda) & 0 \\ 0 & \exp(-i 2 \pi n_l d_l \cos \vartheta_l / \lambda) \end{pmatrix},$$

$l$ stands for the layer index with $l = 0$ and $l = s$ corresponding to the fiber core and the external buffer layer, respectively, $d_l$ and $n_l$ are the thickness and the refractive index of the $l$-th layer, $\vartheta_l$ is the angle of light incidence on the interface between ($l$-1)-th and $l$-th layers. In the core and all layers filled by air, one can derive $\cos \vartheta_l = \lambda / (2 n_0 d_0)$ for the grazing light incidence. Hence, it is easy to obtain from the Snell's law that $\cos\vartheta_l = \sqrt{1-(n_0/n_l)^2}$ in the glass and nanocomposite bilayers. In the analysis, we neglect the buffer PDDA layers since their thickness is around several nanometers and their refractive index is close to the refractive index of the glass.

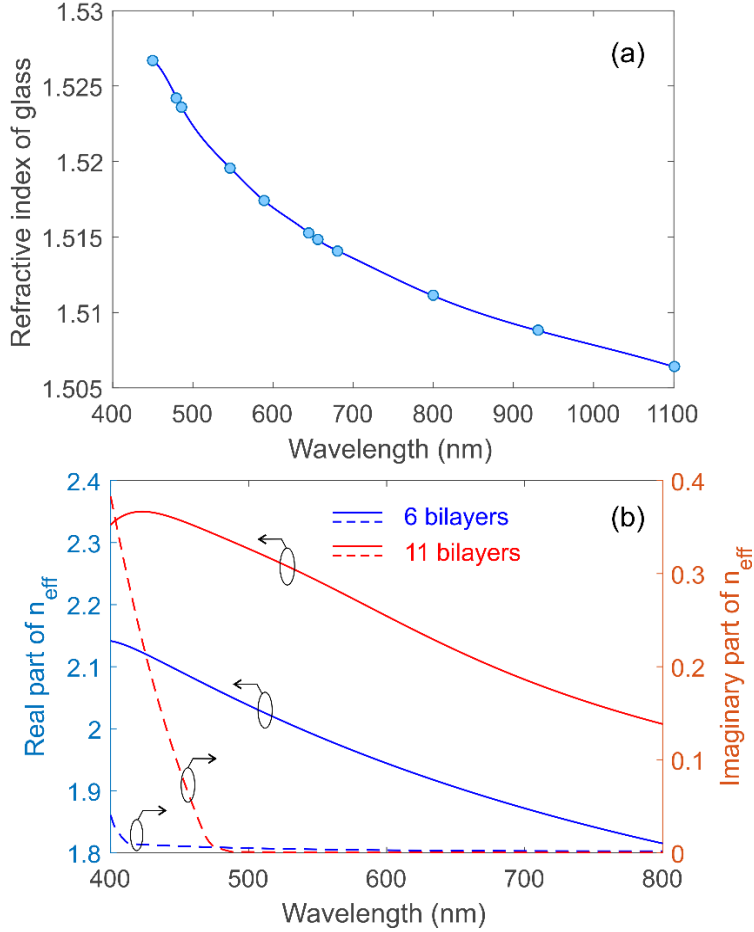

**Fig. 5.** Dependence of the refractive index for (a) the custom-made glass and (b) composite bilayers with MNPs on the light wavelength. In (a) points denote measurements, and the curve is an interpolation. The dispersion of the effective refractive index for the composite bilayers has been adopted from [13].

Finally, we use the measured dispersion of the refractive indexes for the fiber glass (Fig. 5 (a)) and nanocomposite layers (Fig. 5 (b)) [13] to calculate the HC-MF transmission in the visible and near infra-red spectral domains by substituting Eqs. (2) and (3) in Eq. (1) for TE-polarization. Specifically, the transmission windows of the unfilled HC-MF totally fit the measurements (Fig. 4(a)). In the assumption that the thickness of every nanocomposite bilayer is 25 nm, we also demonstrate an excellent agreement with the experiment for the spectral shift in the transmission induced by the nanocomposite coating (Fig. 4(c)). The estimation of the thickness for a single nanocomposite bilayer is close to the sum of the measured nanoparticle size (18 ± 5 nm, see Appendix B) and the thickness of a PDDA layer (1.5 ± 0.5 nm). The slightly larger theoretical estimation for a bilayer thickness can be attributed to aggregation of MNPs during the coating formation. Although in the experiment the light polarization has not been controlled, the agreement with the theoretical predictions obtained for TE-polarization allows us to believe that TM-polarization was much less effectively coupled to the fiber.

It is instructive to note that the nanocomposite coating by more than 5 bilayers leads to significant optical losses due to increasing in the absorption of the nanocomposite bilayers, as shown in Fig. 5 (b) [13]. Moreover, it has been demonstrated that the average coating roughness is increasing with the growth in the number of layers [12,32]. This stems from aggregation of inorganic nanoparticles into clusters, whose quantity and size are directly related with the number of layers. Thus, the too large number of bilayers not only results in artifacts in MRI scans but also deteriorates the transmission properties of HC-MFs.

Additionally, axial inhomogeneities can appear in the layer thickness since we applied the pipette dispenser for filling HC-MF which does not allow keeping the flow rate constant. As a result, diffusion leads to the undesirable gradient in concentration of the applied solution. This issue, however, can be resolved by using a microfluidic pump unit, providing continues and highly controllable flow rate.

## 3. Conclusion and outlook

In conclusion, we demonstrated the novel approach to make optical-fiber-based endoscopes visible in MRI. The high-contrast MRI scans of the HC-MFs have been obtained via Layer-by-Layer assembly of bilayers containing poly(diallyldimethylammonium) chloride and magnetite nanoparticles on the inner surfaces of the fiber core and the cladding capillaries. We found that the optimal number of bilayers to avoid artifacts in MRI and increasing fiber losses should be less than 5. The excellent agreement for the fiber transmission spectra between measurements and theoretical calculations has been demonstrated.

Our findings can be transferred to a variety of hollow-core microstructured and photonic-crystal fibers, providing great capacity for practical use. Specifically, the use of endoscopic optical bioimaging together with MRI has the potential to considerably improve a broad range of diagnostic and therapeutic medical procedures. This approach would enable accurate bioimaging of the area of interest immediately after insertion of the endoscopic probe without unnecessary damage of living tissue that would result in shortening of a postoperative recovery period. High-contrast probe visualization in real time is a must for the study of neuronal activity in deep brain areas *in vivo* since the poorly-arranged probe path may seriously affect the results. Beyond that, our technique can be used as a new tool for post-processing spectral tuning of fiber transmission windows that can find applications in fiber optics.

## Appendix A: Characterization of refractive index for custom-made soft glass used for drawing of fiber samples

The measurement of the dependence of refractive index (RI) of the glass on the radiation wavelength (Figure 5(a)) was carried out by using the multi-wavelength Abbe refractometer DR-M2/1550 (Atago, Japan). Multi-wavelength refractometer Abbe allows one to measure the RI in the wavelength range 450-1550 nm with an accuracy of ± 0.0002. As the source of radiation, we used a high-power candescent lamp. The wavelength of light is determined by the selection of the particular interferential filter. Available interferential filters allowed for measurements on the wavelengths 450±2 nm, 480±2 nm, 486±2 nm, 546±2 nm, 589±2 nm, 644±2 nm, and 656±2 nm, 680±5 nm, 800±5 nm, 930±6 nm, 1100±26 nm, 1300±25 nm 1550±25 nm. The calibration of the device by measuring RI of the prism (n=1.5161) at the wavelength of 589 nm (the absorption band of sodium) was used at the beginning of experiment. Mono bromonaphthalene was used as a contact liquid. The average measurement error of the RI was ±0.0003. The prism temperature during the measurements was kept being +21.5°C by means of water circulation in the refractometer.

## Appendix B: Characterization of PDDA and magnetite nanoparticles

PDDA of different average MW (low 100-200 kDa, medium 250-350 and high 400-500 kDa) with concentration 2 mg/ml in aqueous 0,15 M NaCl solution and a hydrosol of MNPs were used for modification of the HC-MFs internal structure. MNPs were synthesized by using the method described by German at al [33]. and stabilized by citric acid. Finally, the magnetite nanoparticles had average size 18±5 nm and zeta-potential −32±8 mV (Figure 6). In all experiments, deionized water was used, purified by a three-stage Millipore Milli-Q Plus 185 purification system, and had a resistivity higher than 18.2 MΩ cm. TEM image of the magnetite nanoparticles was obtained with a Libra-120 transmission electron microscope (Carl Zeiss, Germany) operating at 120 kV. The magnetite nanoparticles were deposited on a 300-mesh copper grid coated with formvar.

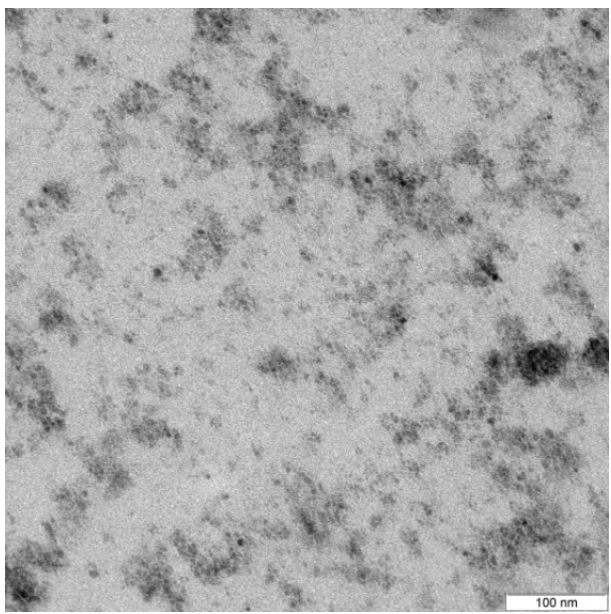

**Fig. 6.** Image of magnetite nanoparticles obtained by transmission electron microscopy (TEM).

## Funding

Russian Foundation for Basic Research RFBR, No. 18-29-08046; ERC StG 'In Motion', PAZY Foundation (Grant No. 01021248), Tel Aviv University Breakthrough Innovative Research Grant.

## Acknowledgement

The authors thank Dr. Andrey M. Zaharevich and MD Victor V. Zuev for SEM and MRI measurements, respectively.

## References

1. X. D. Li, S. A. Boppart, J. Van Dam, H. Mashimo, M. Mutinga, W. Drexler, M. Klein, C. Pitris, M. L. Krinsky, M. E. Brezinski, and J. G. Fujimoto, "Optical coherence tomography: advanced technology for the endoscopic imaging of Barrett's esophagus," Endoscopy **32**(12), 921–930 (2000).
2. T. Kume, T. Akasaka, T. Kawamoto, Y. Ogasawara, N. Watanabe, E. Toyota, Y. Neishi, R. Sukmawan, Y. Sadahira, and K. Yoshida, "Assessment of coronary arterial thrombus by optical coherence tomography," Am. J. Cardiol. **97**(12), 1713–1717 (2006).
3. S. Lam, B. Standish, C. Baldwin, A. McWilliams, J. leRiche, A. Gazdar, A. I. Vitkin, V. Yang, N. Ikeda, and C. MacAulay, "In vivo optical coherence tomography imaging of preinvasive bronchial lesions," Clin. Cancer Res. **14**(7), 2006–2011 (2008).
4. B. A. Flusberg, E. D. Cocker, W. Piyawattanametha, J. C. Jung, E. L. M. Cheung, and M. J. Schnitzer, "Fiber-

optic fluorescence imaging," Nat. Methods **2**(12), 941–950 (2005).
5. K.-B. Sung, C. Liang, M. Descour, T. Collier, M. Follen, and R. Richards-Kortum, "Fiber-optic confocal reflectance microscope with miniature objective for in vivo imaging of human tissues," IEEE Trans. Biomed. Eng. **49**(10), 1168–1172 (2002).
6. B. A. Flusberg, A. Nimmerjahn, E. D. Cocker, E. A. Mukamel, R. P. J. Barretto, T. H. Ko, L. D. Burns, J. C. Jung, and M. J. Schnitzer, "High-speed, miniaturized fluorescence microscopy in freely moving mice," Nat. Methods **5**(11), 935–938 (2008).
7. H. Pahlevaninezhad, M. Khorasaninejad, Y.-W. Huang, Z. Shi, L. P. Hariri, D. C. Adams, V. Ding, A. Zhu, C.-W. Qiu, F. Capasso, and M. J. Suter, "Nano-optic endoscope for high-resolution optical coherence tomography in vivo," Nat. Photonics **12**(9), 540–547 (2018).
8. T. Čižmár and K. Dholakia, "Exploiting multimode waveguides for pure fibre-based imaging," Nat. Commun. **3**(1), 1027 (2012).
9. S. Turtaev, I. T. Leite, T. Altwegg-Boussac, J. M. P. Pakan, N. L. Rochefort, and T. Čižmár, "High-fidelity multimode fibre-based endoscopy for deep brain in vivo imaging," Light Sci. Appl. **7**(1), 92 (2018).
10. R. K. Iler, "Multilayers of colloidal particles," J. Colloid Interface Sci. **21**(6), 569–594 (1966).
11. G. Decher, J. D. Hong, and J. Schmitt, "Buildup of ultrathin multilayer films by a self-assembly process: III. Consecutively alternating adsorption of anionic and cationic polyelectrolytes on charged surfaces," Thin Solid Films **210–211**, 831–835 (1992).
12. D. Grigoriev, D. Gorin, G. B. Sukhorukov, A. Yashchenok, E. Maltseva, and H. Möhwald, "Polyelectrolyte/magnetite Nanoparticle Multilayers: Preparation and Structure Characterization," Langmuir **23**(24), 12388–12396 (2007).
13. A. M. Yashchenok, D. A. Gorin, M. Badylevich, A. A. Serdobintsev, M. Bedard, Y. G. Fedorenko, G. B. Khomutov, D. O. Grigoriev, and H. Möhwald, "Impact of magnetite nanoparticle incorporation on optical and electrical properties of nanocomposite LbL assemblies," Phys. Chem. Chem. Phys. **12**(35), 10469–10475 (2010).
14. I. Dincer, O. Tozkoparan, S. V. German, A. V. Markin, O. Yildirim, G. B. Khomutov, D. A. Gorin, S. B. Venig, and Y. Elerman, "Effect of the number of iron oxide nanoparticle layers on the magnetic properties of nanocomposite LbL assemblies," J. Magn. Magn. Mater. **324**(19), 2958–2963 (2012).
15. M. Balakrishnan, R. Spittel, M. Becker, M. Rothhardt, A. Schwuchow, J. Kobelke, K. Schuster, and H. Bartelt, "Polymer-Filled Silica Fibers as a Step Towards Electro-Optically Tunable Fiber Devices," J. Lit. Technol. **30**(12), 1931–1936 (2012).
16. R. Spittel, D. Hoh, S. Brückner, A. Schwuchow, K. Schuster, J. Kobelke, and H. Bartelt, "Selective filling of metals into photonic crystal fibers," in *Proc. SPIE*, A. Adibi, S.-Y. Lin, and A. Scherer, eds. (2011), **7946**, p. 79460Z.
17. A. Hassani and M. Skorobogatiy, "Design of the microstructured optical fiber-based surface plasmon resonance sensors with enhanced microfluidics," Opt. Express **14**(24), 11616–11621 (2006).
18. A. Hassani and M. Skorobogatiy, "Photonic crystal fiber-based plasmonic sensors for the detection of biolayer thickness," J. Opt. Soc. Am. B **26**(8), 1550 (2009).
19. M. A. Schmidt, N. Granzow, N. Da, M. Peng, L. Wondraczek, and P. S. J. Russell, "All-solid bandgap guiding in tellurite-filled silica photonic crystal fibers," Opt. Lett. **34**(13), 1946–1948 (2009).
20. D. Lopez-Torres, C. Elosua, J. Villatoro, J. Zubia, M. Rothhardt, K. Schuster, and F. J. Arregui, "Photonic crystal fiber interferometer coated with a PAH/PAA nanolayer as humidity sensor," Sens. Actuators B Chem. **242**, 1065–1072 (2017).
21. J. Yin, S. Ruan, T. Liu, J. Jiang, S. Wang, H. Wei, and P. Yan, "All-fiber-optic vector magnetometer based on nano-magnetic fluids filled double-clad photonic crystal fiber," Sens. Actuators B Chem. **238**, 518–524 (2017).
22. A. A. Chibrova, A. A. Shuvalov, Y. S. Skibina, P. S. Pidenko, S. A. Pidenko, N. A. Burmistrova, and I. Y. Goryacheva, "The red shift of the semiconductor quantum dots luminescence maximum in the hollow core photonic crystal fibers," Opt. Mater. (Amst) **73**, 423–427 (2017).
23. P. Gross, R. I. Kitney, S. Claesen, and J. M. Halls, "MR-compatible endoscopy and tracking for image-guided surgery," Int. Congr. Ser. **1230**, 1076–1082 (2001).
24. J. S. Skibina, R. Iliew, J. Bethge, M. Bock, D. Fischer, V. I. Beloglasov, R. Wedell, and G. Steinmeyer, "A chirped photonic-crystal fibre," Nat. Photonics **2**(11), 679–683 (2008).
25. Y. S. Skibina, V. V. Tuchin, V. I. Beloglazov, G. Shteinmaeer, I. L. Betge, R. Wedell, and N. Langhoff, "Photonic crystal fibres in biomedical investigations," Quantum Electron. **41**(4), 284–301 (2011).
26. J. S. Skibina, A. V. Malinin, A. A. Zanishevskaya, and V. V. Tuchin, "Photonic Crystal Waveguide Sensing," in *Portable Biosensing of Food Toxicants and Environmental Pollutants*, G.-P. N. D. P. Nikolelis, T. Varzakas, A. Erdem, ed. (CRC Press, 2013), 1–32.
27. A. V. Malinin, J. S. Skibina, V. V. Tuchin, M. V. Chainikov, V. I. Beloglazov, I. Y. Silokhin, A. A. Zanishevskaya, V. A. Dubrovskii, and A. A. Dolmashkin, "The use of hollow-core photonic crystal fibres as biological sensors," Quantum Electron. **41**(4), 302–307 (2011).
28. A. A. Zanishevskaya, A. A. Shuvalov, Y. S. Skibina, and V. V. Tuchin, "Blood typing using microstructured waveguide smart cuvette," J. Biomed. Opt. **20**(4), 040503 (2015).
29. N. M. Litchinitser, A. K. Abeeluck, C. Headley, and B. J. Eggleton, "Antiresonant reflecting photonic crystal optical waveguides," Opt. Lett. **27**(18), 1592–1594 (2002).
30. A. M. Zheltikov, "Colors of thin films, antiresonance phenomena in optical systems, and the limiting loss of modes in hollow optical waveguides," Phys. Uspekhi **51**(6), 591 (2008).

31. P. Yeh, *Optical Waves in Layered Media* (Wiley-Interscience, 2005), **61**.
32. A. M. Yashchenok, D. A. Gorin, M. Badylevich, A. A. Serdobintsev, M. Bedard, Y. G. Fedorenko, G. B. Khomutov, D. O. Grigoriev, and H. Möhwald, "Impact of magnetite nanoparticle incorporation on optical and electrical properties of nanocomposite LbL assemblies," Phys. Chem. Chem. Phys. **12**(35), 10469–10475 (2010).
33. S. V. German, O. A. Inozemtseva, A. V. Markin, K. Metvalli, G. B. Khomutov, and D. A. Gorin, "Synthesis of magnetite hydrosols in inert atmosphere," Colloid J. **75**(4), 483–486 (2013).